\documentclass[12pt]{article}   
\usepackage{psfig,times,fancyheadings}  

\begin{document}
\thispagestyle{empty}

 \lhead[\fancyplain{}{\sl }]{\fancyplain{}{\sl }}
 \rhead[\fancyplain{}{\sl }]{\fancyplain{}{\sl }}

 \renewcommand{\topfraction}{.99}      
 \renewcommand{\bottomfraction}{.99} 
 \renewcommand{\textfraction}{.0}


\newcommand{\nc}{\newcommand}

\nc{\qI}[1]{\section{{#1}}}
\nc{\qA}[1]{\subsection{{#1}}}
\nc{\qun}[1]{\subsubsection{{#1}}}
\nc{\qa}[1]{\paragraph{{#1}}}

\def\qbu{\hfill \par \hskip 6mm $ \bullet $ \hskip 2mm}
\def\qee#1{\hfill \par \hskip 6mm #1 \hskip 2 mm}

\nc{\qfoot}[1]{\footnote{{#1}}}
\def\qL{\hfill \break}
\def\qpar{\vskip 2mm plus 0.2mm minus 0.2mm}
\def\qtvi{\vrule height 2pt depth 5pt width 0pt}
\def\qth{\vrule height 12pt depth 0pt width 0pt}
\def\qtb{\vrule height 0pt depth 5pt width 0pt}
\def\tvi{\vrule height 12pt depth 5pt width 0pt}

\def\qparr{ \vskip 1.0mm plus 0.2mm minus 0.2mm \hangindent=10mm
\hangafter=1}

\def\qdec#1{\par {\leftskip=2cm {#1} \par}}

\def\qdpt{\partial_t}
\def\qdpx{\partial_x}
\def\qddpt{\partial^{2}_{t^2}}
\def\qddpx{\partial^{2}_{x^2}}
\def\qn#1{\eqno \hbox{(#1)}}
\def\qds{\displaystyle}
\def\qw{\widetilde}
\def\qmax{\mathop{\rm Max}}   
\def\qmin{\mathop{\rm Min}}   

\def\qb#1{\hbox{#1}}
\def\qd{^{\ \circ}\hbox{C}}
\def\qdk{^{\ \circ}\hbox{K}}
\def\ql{\ \hbox{--}\ }
\def\qh#1#2{\buildrel{\displaystyle #2} \over {#1}}


\def\qci#1{\parindent=0mm \par \small \parshape=1 1cm 15cm  #1 \par
               \normalsize}

\null

\centerline{\bf \Large A bridge between liquids and socio-economic systems:}
\vskip 5mm
\centerline{\bf \Large the key role of interaction strengths}                                      

\vskip 1cm
\centerline{\bf Bertrand M. Roehner $ ^1 $ }
\vskip 4mm
         
\centerline{\bf Institute for Theoretical and High Energy Physics}
\centerline{\bf University Paris 7 }

\vskip 2cm

{\bf Abstract}\quad One distinctive and pervasive aspect of social
systems is the fact that they comprise several kinds of agents. 
Thus, in order to
draw parallels with physical systems one is lead
to consider binary (or multi-component) compounds. 
Recent views about the mixing of liquids in 
solutions gained from neutron and X-ray scattering show 
these systems to have a number
of similarities with socio-economic systems.  
It appears that such phenomena as rearrangement
of bonds in a solution, gas condensation, 
selective evaporation of molecules can 
be transposed in a natural way to socio-economic phenomena.
These connections provide a
novel perspective for looking at social systems which we illustrate 
through some examples. For instance, we
interpret suicide as an escape phenomenon and in order to test that
interpretation we consider social systems characterized by very low
levels of social interaction. For those systems suicide rates are
found to be 10 to 100 times higher than in the general population. 
Another interesting parallel concerns the
phase transition which occurs when locusts gather together to form
swarms which may contain several billion insects. 
What hinders the thorough investigation of such cases 
from the standpoint of collective phenomena
that we advocate is the lack or inadequacy of statistical data for,
up to now, they were collected for completely different purposes.
Most essential for further progress 
are statistics which would permit
to estimate the strength of social ties and interactions. 
Once adequate data become available, rapid advance may be
expected.

\vskip 1cm

\centerline{May 8, 2004}

\vskip 8mm
\centerline{\it Preliminary version, comments are welcome}

\vskip 2cm

1: ROEHNER@LPTHE.JUSSIEU.FR
\qL
\phantom{1: } Postal address where correspondence should be sent:
\qL
\phantom{1: }B. Roehner, LPTHE, University Paris 7, 2 place Jussieu, 
F-75005 Paris, France.
\qL
\phantom{1: }E-mail: roehner@lpthe.jussieu.fr
\qL
\phantom{1: }FAX: 33 1 44 27 79 90

\vfill \eject

\qI{Introduction}

Over the past decade econophysics and sociophysics have been developed
by theoretical physicists who mainly came from statistical physics.
Recent research in econophysics comprised a large body of
empirical inquiries on topics which so far had been largely
ignored by economists or sociologists. In addition,
a number of theoretical tools developed in polymer physics,
spin glass studies, Ising model simulations or discrete scaling 
were tentatively applied to problems in economics and sociology.
This paper develops the idea that there is a connection between
some of the phenomena studied in statistical physics and
processes which occur in human societies. If persuasive, that 
argument would strongly support the claim (which
is at the core of econophysics) made by physicists 
that the insight they have gained
in studying physical systems can indeed be of value in the social sciences
as well. Apart from this broad contention, our investigation will also
tell us which phenomena are most likely to provide a 
good starting point for studying social systems in a fruitful way.
\qpar

In the course of this paper we will see that it is the liquid state which seems
to provide the best bridge to social systems. This is easy to understand
intuitively. Crystallized solids have a structure whose regularity and symmetries
have no match in social systems. On the other hand gases are characterized
by a complete lack of structure which does not match social networks. With
their non trivial and adaptive intermolecular interactions, 
liquids and more specifically solutions offer a better analog 
to socio-economic systems. Glasses, that is to say solids without crystal 
structure, could also be possible candidates but in the present paper we
restrict our attention to solutions. Subsequently we give other, more
technical, arguments in favor of a parallel between solutions and social
systems. 
\qpar

Unfortunately, the liquid state is probably the less well understood.
It has been suspected for a long time (see for instance Moelwyn-Hughes
1957) that the departure from ideal (or even regular) solution behavior
is due to the formation of complex molecular assemblages even for
non-ionic solutions. This was the central assumption on which 
Dolezalek's theory was based; however it is only in 
recent decades, that neutron and X-ray scattering
as well as infra-red spectroscopy provided a more accurate picture
of such molecular clusters. The new picture which progressively emerged
from these studies gave us an insight into microscopic mechanisms at
molecular level. It is at this level that the parallel with social systems becomes
closer and more natural. That is why, throughout this paper, we 
try to stick to molecular mechanisms and refrain from using such concepts
as entropy, energy or temperature which become meaningful only at 
macroscopic level. So far, these concepts have no clear equivalent in
social systems which, by the way, 
is not surprising for in a similarly way at molecular level
the only notions which make sense are those of
distance and molecular attraction, stretching and
vibrations, molecular assemblage, and so on.
In the first part of this paper I describe some physical phenomena
involving solutions in terms which can be easily transposed to social
systems; in the second, I invite the reader to take 
the plunge and outline some social parallels. 
\qpar

The
paper proceeds as follows. In the second section, I recall that 
the key variable which accounts for a whole range of phenomena
as diverse as boiling temperature, vapor pressure, surface tension,
or viscosity is the strength of the intermolecular interaction. This is
particularly true in the liquid state. That observation is a strong 
incentive to develop methods for measuring the strength of social
ties. Once the prime importance of intermolecular coupling has been
recalled, I describe what I call a paradigm experiment because it provides
so to say a blueprint for future studies of social systems. This and similar
experiments with solutions suggest that seeing the mixing of two liquids
merely from the point of view of entropy as an irreversible operation which
increases disorder prevents us from seeing the major role played
by amalgamating and combining,
two mechanisms which play a key role in 
both biological and social phenomena.
In  section 4, I consider the phenomenon of
suicide in situations in which one has good
reason to expect a low level of social interaction and, accordingly,
high suicide rates. Then
I devote a few words to social or biological situations
which are similar to gas condensation or solvation.
Needless to say, each of these
phenomena would deserve  a more detailed study. Our objective in this
paper is to draw a possible agenda for future research rather than
to offer detailed case-studies. 
\vskip 10mm

\centerline{\bf \Large Part I \quad Physical background}
\vskip 5mm

Studying social phenomena is often frustrating because for each law
or regularity that one tentatively
tries to propose there are usually many exceptions and outliers. 
The situation is fairly similar in physical chemistry.
No model has a broad validity and exceptions abound
even for the most basic effects. In that sense physical chemists are
certainly better prepared to cope with social systems than for instance
particle physicists or solid state physicists.
Our background presentation in the first
part is entirely based on experimental evidence; the main reason for
avoiding theoretical concepts is the fact that they cannot be easily 
adapted to social systems. The discussions I have had with the colleagues
in my lab convinced me that even experienced theoretical physicists may
not necessarily be familiar with those facts and interpretations; however,
this first part may be safely skipped by physical chemists.

\qI{The key role of interaction strengths}

In statistical physics we
know that, at least in principle,
the properties of a system may be derived from
its Hamiltonian. However, for systems like liquids
this can only be done with great difficulty and often only
numerically. The point we want to make in this section is much simpler.
We show that the behavior of a system to a large extent depends only
on the {\it strength} of the interaction; its precise form,
whether it is a coupling between ions, permanent dipoles,
induced dipoles or a mix of those interactions does not really matter
from an experimental point of view.
In what follows the interaction strength will be our key parameter.
It has by the way a simple geometrical interpretation in the sense that
there is a close relationship between interaction strength and
length of intermolecular bonds. 
\qpar

Before looking at more elaborate properties we first consider 
the state, whether solid, liquid or gas, in which a compound is to be
found at room temperature. At first sight our claim that it is determined
by the coupling strength could seem utterly wrong. Compare for instance
$ \hbox{H}_{\hbox{2}}\qb{O} $ and  $ \qb{H}_{\qb{2}}\qb{S} $; while the
first is a liquid, the second is a gas (boiling temperature = $ -60 \qd $)
Yet, these molecules could be expected to be fairly similar for
sulfur and oxygen are in the same column of Mendeleev's periodic table.
Perhaps an even more striking apparent counter-example is ethanol and
methyl ether which both correspond to the formula 
$ \qb{C}_{\qb{2}}\qb{H}_{\qb{6}}\qb{O} $. Whereas the first is a liquid (boiling
point = $ 78 \qd $), the second is a gas (bp=$-24 \qd $). Thanks to the
thorough and patient work of physical chemists, we now know that the 
interactions between $ \hbox{H}_{\hbox{2}}\qb{O} $ molecules is 
indeed much stronger than the one between $ \qb{H}_{\qb{2}}\qb{S} $
molecules and similarly for ethanol and methyl ether molecules. It would
take us too far away from the main purpose of this paper to explain
the reasons of these differences. The main point we want to emphasize
is that, in contrast to mainstream ideas in the 1910s (see for instance
Holmes 1913)
overall properties such a size, mass or composition of molecules
turned out  to be
completely inadequate to explain their properties; that could be done
only once experimental methods had been developed which could provide
reliable estimates of intermolecular coupling. 
\qpar

Building on this knowledge we present in Fig.1a  some data which illustrate
the key role of the coupling strength. 
  \begin{figure}[htb]
    \centerline{\psfig{width=15cm,figure=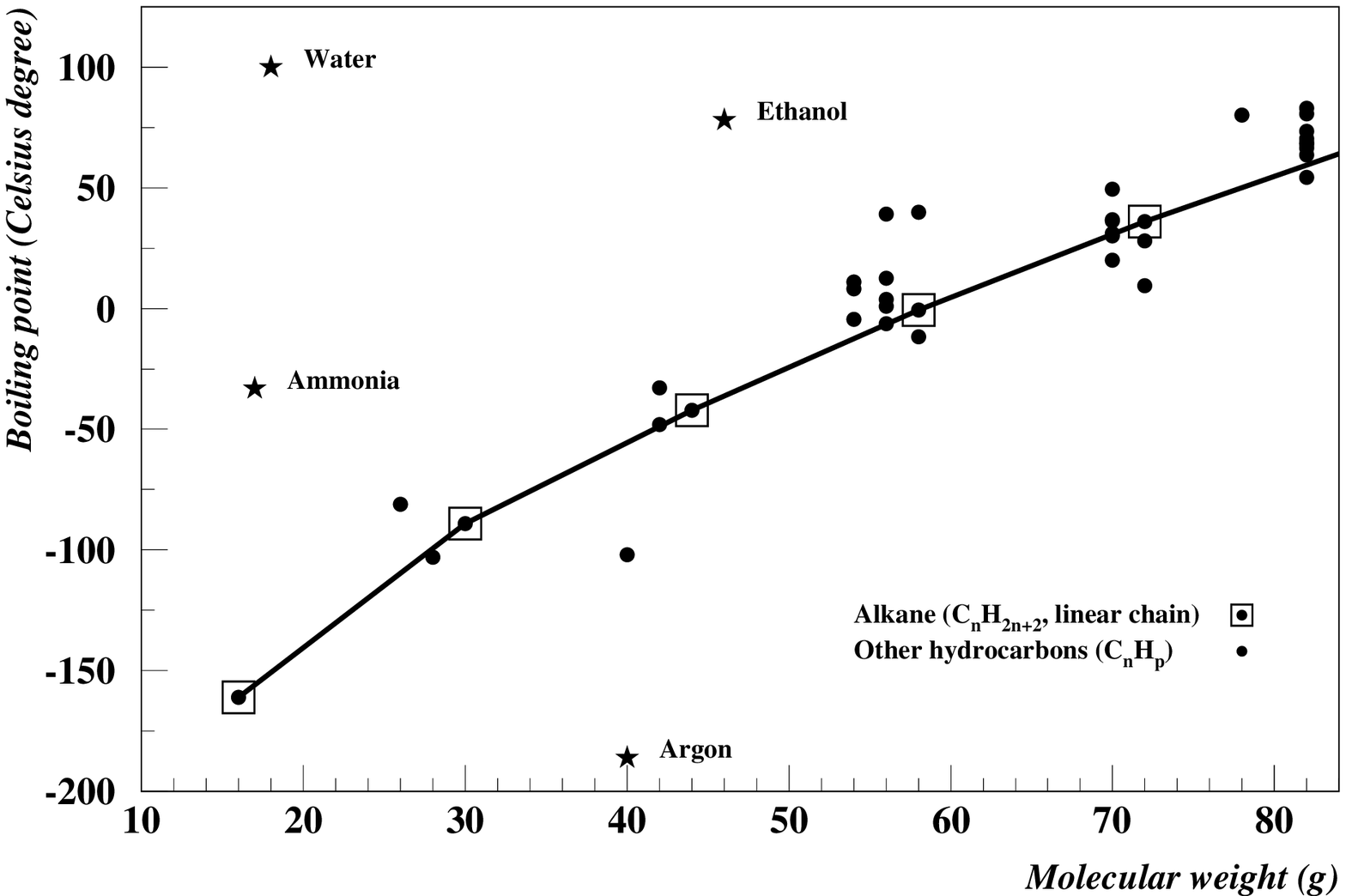}}
    {\bf Fig.1a: Boiling temperature as a function of intermolecular
attraction}.
{\small For alkanes $ \qb{C}_n\qb{H}_{2n+2} $ with a linear chain,
which are represented by dots surrounded by a square, the inter-molecular
attraction is proportional to the number of the hydrogen atoms
and hence
also to the molecular weight $ M=14n+2 $. The trend portrayed by the
solid line means that for longer carbon chains more thermal agitation is
required in order to break the intermolecular bonds. The dots represent
hydrocarbons $ \qb{C}_n\qb{H}_{p} $ whose intermolecular forces,
are slightly different due for instance to  branched carbon chain which
results in boiling temperature differences of the order of 10\%. The stars
correspond to compounds whose molecular coupling are of a different
nature, either much weaker (argon) or much stronger (ammonia, ethanol,
water)}.
{\small \it Source: Lide (2001)}.
 \end{figure}

Apart from the outliers to which we
come back below, the graph concerns only hydrocarbons and more
particularly alkanes: 
$ \qb{C}_n\qb{H}_{2n+2} $. As is well known, alkane molecules
interact only trough dipole-induced forces, the so-called London dispersion
forces. These fairly weak forces exist between any pair of atoms. As a result
the interaction between two alkane molecules is basically proportional
to the length of the carbon chain that is to say to the number $ n $ or in other
words to the molecular weight of the alkane. The square correspond to
experimental data for linear alkane chains, whereas the dots correspond
to other hydrocarbons. Of particular interest are the dots which correspond
to isomers (same molecular weight) of a given alkane. These isomers have
ramified carbon chains, a feature which to some extent changes
the London forces between the molecules and results in
differences of the order of 10\% (when
temperatures are expressed in Kelvin degrees). The stars show a number
of cases characterized by different kinds of interactions. As we know
both water and ethanol, 
$ \qb{C}\qb{H}_{\qb{3}} \ql \qb{C}\qb{H}_{\qb{2}} \ql \qb{O}\qb{H} $
have a dipole $ \qb{O}_{-}\qb{H}_{+} $ which causes a fairly strong interaction
trough so-called hydrogen bonds. At the bottom of the graph the single
atoms of argon have almost no interaction at all which results in a very low
boiling point close to $ -200 \qd $. Incidentally, it can be observed that there
is a close relationship between bond strength and bond length. For instance
if we assume a potential corresponding to a ion-ion attraction and a hard
core repulsive force proportional to $ 1/ r^p,\ (p\sim 9) $ 
the strength $ s $ of the
bond is related to the distance $ R $ between the two ions by the relation:
$$ s=\left. { d^2 V \over dr^2 }\right|_{r=R} 
=  (p-1){ e^2 \over 4\pi \epsilon _0 }{1 \over R^3 } $$

In order to show that the previous argument extends to many other 
physical properties we show in Fig.1b that the enthalpy of vaporization and
the viscosity of alkanes is again determined by the 
strength of the intermolecular forces that is
to say in this case by the molecular weight. As it would
be pointless to compare the viscosity of gases with that of liquids we restricted
the latter curve to the alkanes which are liquid at room temperature. 
\qpar
  \begin{figure}[htb]
    \centerline{\psfig{width=15cm,figure=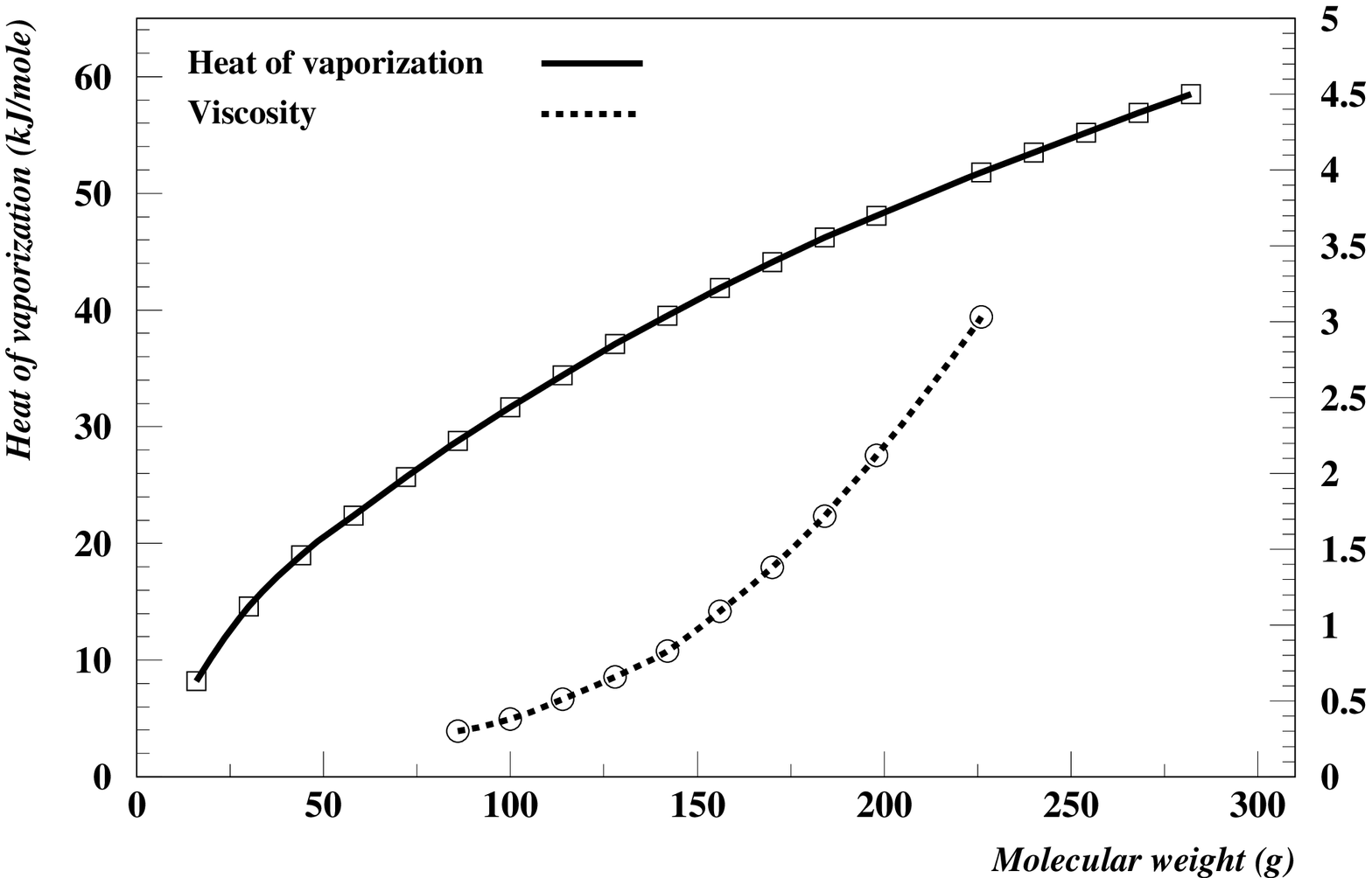}}
    {\bf Fig.1b: Latent heat of vaporization and viscosity as a function of
inter-molecular attraction for alkanes}.
{\small As explained in Fig.1a, there is a direct relationship between
attraction strength and molecular weight. The solid line corresponds to
the 20 first alkanes (except $ n=15 $ which is missing in data tables); it
describes the empirical relationship: $ L_s (\qb{C}_n\qb{H}_{2n+2})=
1.1+1.7 n $.
The broken line represents the viscosity; it is
restricted to the alkanes which are liquid at room
temperature, namely $ n=7,\ldots ,16 $ ($ n=15 $ is again missing)}.
{\small \it Sources: Lide (2001), Moelwyn-Hughes (1961, p.702)}.
 \end{figure}

The relationships displayed by the curves in Fig.1a,b have a clear intuitive
interpretation. The stronger the interaction, the better the molecules are
held together and the more kinetic energy it takes to disrupt the molecular
assemblages that make up solids or liquids. In the same way, in a liquid
with a strong interaction only the fastest molecules will be able to escape
which translates into a low vapor pressure. As to viscosity, a strong 
interaction will make neighboring layers to stick more closely together 
which for liquids results in a higher viscosity. 
\qpar

In order to test this 
way of reasoning let us see if we can use it in order to predict the
relationship between interaction strength and other physical properties
such as for instance the speed of sound.
For sound to propagate, successive layers must be put into motion.
Due to inertia,
in order to put one layer into motion the main factor is the weight of the
molecule. If there are strong bonds between molecules two things 
will happen.
\qbu In a given layer, the inertia effect will be increased because 
strongly coupled molecules will somewhat behave as a cluster of
molecules that is to say a super molecule of greater molecular weight.
\qbu The transmission of the perturbation from one layer to the next
will be facilitated. 
\qpar

Which one of these effects will prevail is not obvious. Observation shows
that the first effect prevails in gases. Thus, by comparing the speed of
sound in methane, $ \qb{C}\qb{H}_{\qb{4}} $, and propane,
$ \qb{C}_{\qb{3}}\qb{H}_{\qb{8}} $, we see that it is smaller in propane
even after the factor $ \sqrt{M} $ due to the molecular weight has been
corrected for. 
In other words, for gases the speed of sound decreases with 
stronger interactions.
On the contrary, in liquids and solids it is the second
effect which prevails as illustrated by the two following examples%
\qfoot{Let us recall that for all materials, whether gas, liquid or solid,
the speed of sound is given by $ c=\sqrt{E/\rho} $ where 
$ \rho $ is the density and $ E $ the
bulk modulus of elasticity $ E=\Delta \hbox{pressure}/[\Delta \rho/\rho ] $
which basically describes the hardness of the
material. 
In the case of a gas $ E $ can be expressed
as the inverse of the (adiabatic) compressibility $ E=1/[(\Delta V/V)/p] $
where $ V $ and $ p $ denote volume and pressure respectively. In the
case of a solid $ E $ is usually referred to as Young's modulus. In short, the
above
formula says that the speed of sound is larger in harder or lighter
materials.}%
.
(i) The speed of sound in pentane, $ \qb{C}_{\qb{5}}\qb{H}_{\qb{12}} $
is 1012 m/s (at $ 25 \qd $ and 1 bar); thus, on account of its higher
molecular mass one would expect a smaller velocity for heptane,
$ \qb{C}_{\qb{7}}\qb{H}_{\qb{16}} $, yet it is higher at 1129 m/s.
(ii) As one knows, diamond, a solid with very strong interatomic bonds
has a velocity of sound of 12,000 m/s, one of the highest to be observed
in any substance. 
\qpar

The previous discussion shows that even in cases where the physical
consequences of a strong interaction are less transparent than in
the cases of Fig.1a,b, this factor nevertheless plays an essential role.
In the next section we examine the role of interaction strengths in 
the mixing of two liquids. 

\qI{The mixing of two liquids}

The mixing of two liquids is described in Fig.2a. 
%
  \begin{figure}[htb]
    \centerline{\psfig{width=8cm,figure=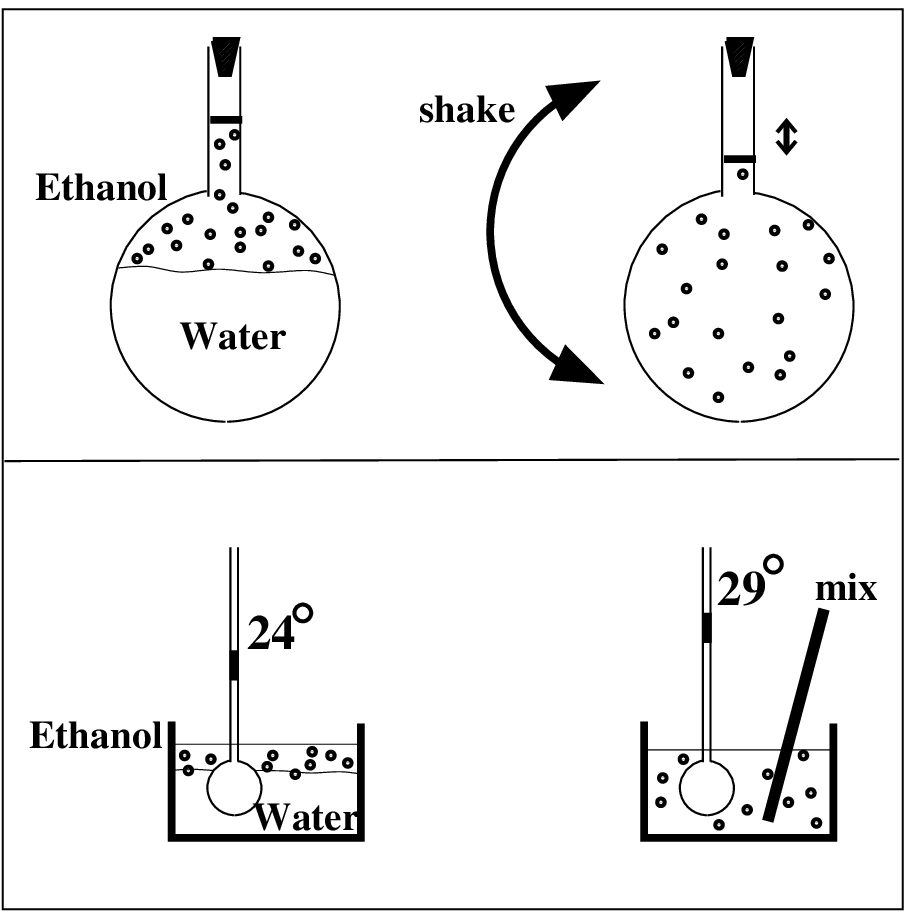}}
\vskip 5mm
    {\bf Fig.2a: Two experiments on the mixing of water and ethanol.}
{\small Not just for fun \ldots
These are not {\it Gedanken} experiments, they can be
performed easily by any theoretical physicist who wants to get a more
intuitive feeling of the mechanisms described in this paper.
When ethanol is added to water there is both a volume contraction and
a release of heat. Both phenomena are affected by respective proportion
and temperature. As predicted by LeChatelier's principle, the effect
of a temperature decrease is to increase the heat of mixing. Similar
experiments can be performed with many other compounds. 
The mixing of acetone ($ \qb{C}_{\qb{3}}\qb{H}_{\qb{6}}\qb{O} $) 
and chloroform ($ \qb{C}\qb{H}\qb{Cl}_{\qb{3}} $) is even more exothermic,
whereas the mixing of acetone 
with carbon disulphide ($ \qb{C}\qb{S}_{\qb{2}} $) is strongly endothermic
and results in a volume increase. Note that because
carbon disulphide is more
toxic and dangerous to handle (it catches fire very easily) than the previous
compounds, the last experiment should rather be done in a chemistry lab.}
{\small \it Source: Moelwyn-Hughes (1961, p. 812)}.
 \end{figure}
%
To call this a paradigm
experiment could seem an inflated expression for such a modest
experiment. Nevertheless, we will see that it carries a number of 
important ideas. Accurate experiments
of this kind were carried out in the early 20th century in particular by
the German physicist Emil Bose (1907). The experiment described in
Fig.2a does not aim at precision, its objective is rather to give an intuitive
feeling of the phenomenon. The upper line in Fig.2a shows that after being
mixed the volume of a solution of water and ethanol,
$ \qb{C}\qb{H}_{\qb{3}} \ql \qb{C}\qb{H}_{\qb{2}} \ql \qb{O}\qb{H} $
decreases. The 
experiment provides a rough estimate of the contraction which is of the order
of 2\%. The lower line shows that the mixing phenomenon is exothermic.
Naturally, the quantity of heat which is released would not be the same
for another alcohol such as methanol,
$ \qb{C}\qb{H}_{\qb{3}} \ql \qb{O}\qb{H} $,
or propanol,
$ \qb{C}\qb{H}_{\qb{3}} \ql \qb{C}\qb{H}_{\qb{2}} \ql \qb{C}\qb{H}_{\qb{2}} 
\ql \qb{O}\qb{H} $. A more detailed picture is given in Fig 2b. 
%
  \begin{figure}[htb]
    \centerline{\psfig{width=15cm,figure=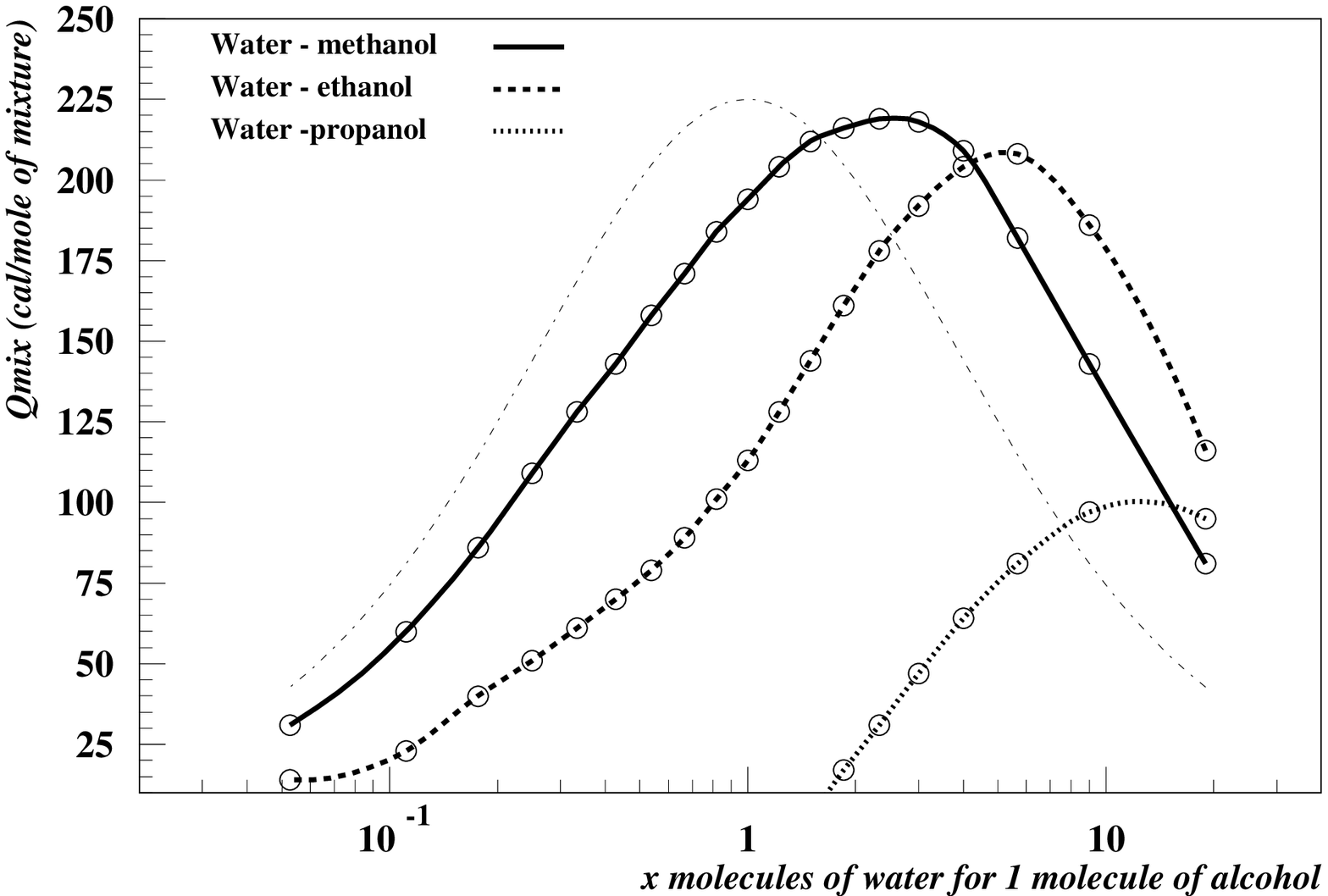}}
    {\bf Fig. 2b: Heats of mixing: water - alcohols}.
{\small Methanol, ethanol and propanol are the first three alcohols:
$ \qb{C}_n\qb{H}_{2n+1}\qb{OH},\ n=1,2,3 $. Usually,
for instance for acetone-chloroform or methanol-ethanol,
the corresponding curves are symmetrical with respect to molar
concentration (as indicated by the thin line curve); this points to a
a connection between the shape of the curves and the
structure of molecular assemblages: the dissymmetry shows that
several water molecules surround each alcohol molecule.
Note that when the proportion
of water becomes too small, the mixing with propanol becomes endothermic.}
{\small \it Source: Bose (1907), Landolt-B\"ornstein (1976).}.
 \end{figure}
Bose's 
results furthermore show that when the experiment is carried out at
$ 43 \qd $ the mixing becomes endothermic as soon as the molar proportion
of propanol becomes higher than 15\%. 
\qpar

Having presented the facts, let us now see what can be learned from
them and why this experiment is of interest for social phenomena.
We will proceed from macroscopic to microscopic level.
\qee{1)} From what has been said in section 1, it is obvious that
the contraction and temperature increase are related. The contraction
shows that intermolecular attraction in the solution is on average 
stronger than in the pure compounds. As the molecules rearrange
themselves in line with the new interactions, energy is released in the
same way as when an expanded spring returns to its equilibrium
length%
\qfoot{Yet, there are cases where exothermic mixing is accompanied by
a dilatation instead of a contraction, for instance in 
in the mixing of ethyl acetate,
$ \qb{C}\qb{H}_{\qb{3}} \ql 
\qh{\qb{C}}{\qh{\parallel}{\qb{O}}}
\ql \qb{O} \ql \qb{C}\qb{H}_{\qb{2}} 
\ql \qb{C}\qb{H}_{\qb{3}} $, 
and carbon disulphide, $ \qb{S} = \qb{C} = \qb{S} $. It is true that the
dilatation is fairly small (only 0.03\%) but this exception nevertheless 
illustrates our previous reflexion about the difficulty in this field
of stating rules which hold without any exception.}%
.
\qee{2)} What has thermodynamic to say about the mixing of liquids?
Because the mixing is exothermic we know that the solution
is not an ideal solution. The heat of mixing is given in terms of
partial pressures by a formula first proposed
by Nernst (Bose 1907, p. 621):
$$ Q_{\hbox{mix}} = -RT^2 { d \over dT} \left[ x\log{ p_a \over p'_a }
+ (1-x)\log{p_b \over p'_b }  \right] \qn{3.1} $$

where $ T $ denotes the Kelvin temperature, $ x $ the molar proportion
of water, $ p_a,\  p'_a $ the pressure of vapor over pure water and
over the solution respectively and $ p_b,\  p'_b $ the same partial pressures
for alcohol. When Raoult's law applies $ p'_a = xp_a $ and similarly
$ p'_b = (1-x)p_b $ which means that the terms between square 
brackets become independent of $ T $ and, as a result, 
$ Q_{\hbox{mix}} =0 $. This is the ideal solution case. 
For non-ideal solutions, the application of the above formula requires
detailed input information about how the partial pressures depend upon
temperature. 
\qee{3)} An interesting question is to understand the location of the peaks
in Fig.2b. Why, for instance is $ Q_{\hbox{mix}} $ maximum for a proportion
of 5 molecules of water for one molecule of ethanol? This leads us to
examine the phenomenon at molecular level.
\qee{4)} Thanks to the extensive work done by physical chemists we now have
a better understanding of the mixing of liquids at molecular level. Of cardinal
importance in the case of ethanol and water is the fact that the ethanol
molecule $ \qb{C}\qb{H}_{\qb{3}} \ql \qb{C}\qb{H}_{\qb{2}} \ql \qb{O}\qb{H} $
comprises two sections which react to water molecules in very
different ways. The $ \qb{OH} $ segment holds a dipole $ \qb{O(-)H(+)} $
which can link up with the $ \qb{O(-)H(+)} $ dipoles of the water molecules.
The segment $ \qb{C}\qb{H}_{\qb{3}} \ql \qb{C}\qb{H}_{\qb{2}} $, on the
other hand, is similar to ethane 
$ \qb{C}\qb{H}_{\qb{3}} \ql \qb{C}\qb{H}_{\qb{3}} $
and we know that ethane like any alkane is miscible in water in very
small proportion only. As a result, this segment could seem to be irrelevant
as far as the mixing with water is concerned. This, however, would be 
a simplistic view. Neutron and X-ray scattering experiments have  shown
that water molecules form a kind of net around alkane molecules in
solution (Schmid 2001, Baumert et al. 2003)%
\qfoot{Such a structure is often referred to as a clathrate, a synonym to
hydrate which comes from the oil industry and designates compounds formed
of ice in which molecules of hydrocarbons are trapped. Such compounds 
form in pipelines which cross cold regions and are also assumed to
exist on some of the satellites of Jupiter or Saturn.}%
. 
The molecular assemblage between water and ethanol molecules is
described schematically in Fig. 3. 
%
  \begin{figure}[htb]
    \centerline{\psfig{width=8cm,figure=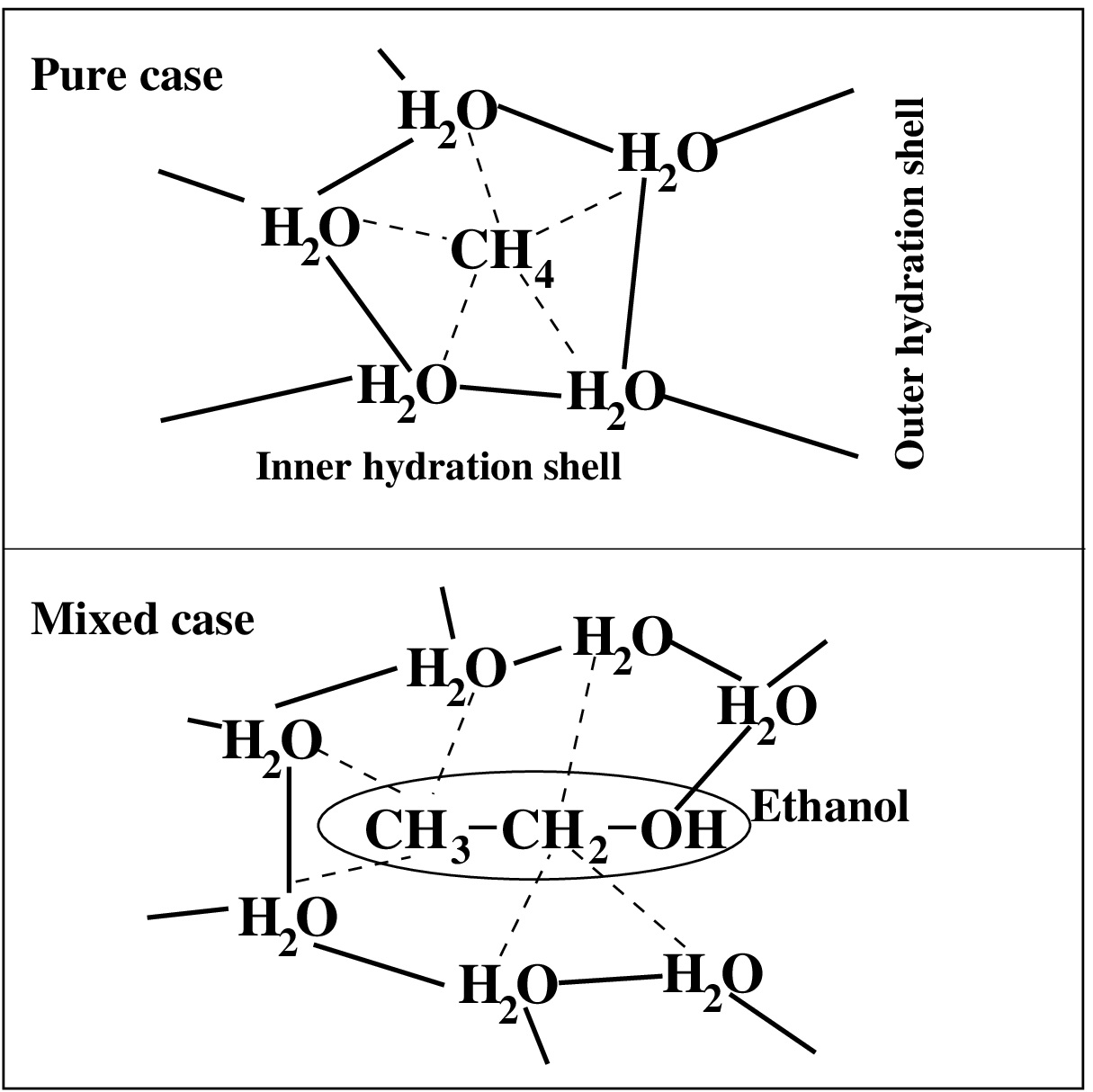}}
\vskip 5mm
    {\bf Fig. 3: Schematic representation of the molecular assemblage in
water-methane and water-ethanol solutions}.
{\small In contrast to methane $ \qb{C}\qb{H}_{\qb{4}} $ which
features only weak London dispersion forces, the molecule of ethanol
comprises two segments 
(i) the alkane-like segment $ \qb{C}\qb{H}_{\qb{3}} \ql \qb{C}\qb{H}_{\qb{2}} $
(ii) the water-like end $ \qb{OH} $. For that reason, ethanol displays
a dual behavior: like methane, it attracts an hydration shell of water 
and like water it forms strong hydrogen bonds. According to some
recent studies (Dill et al. 2003, p. 581) there may be as many as 17 water
molecules in the first hydration shell. The precise shape of the molecular
assemblage is of little importance for the purpose of this paper; what matters
is the fact that it is a highly ordered arrangement}.
{\small \it Sources: Baumert et al (2003), Dill et al. (2003),
Dixit et al. (2002), Guo et al. (2003), Israelachvili et al. (1996)}.
 \end{figure}
%
The solid lines represent the dipole-dipole
bonds while the dashed lines represent the weaker bonds between 
induced dipoles. One should bear in mind that these bonds and indeed
the whole assemblage are rearranged on a picosecond time scale. 
In other words any representation such as the one in Fig. 3 can be nothing
but an average view of a rapidly changing structure. 
In dynamical terms the mixing can be described by the following reaction:
$$ x\qb{H}_{\qb{2}}\qb{O} + y\qb{C}_{\qb{2}}\qb{H}_{\qb{5}}\qb{O} \
\qh{\longrightarrow}{\longleftarrow} \
\left( x\qb{H}_{\qb{2}}\qb{O}, y\qb{C}_{\qb{2}}\qb{H}_{\qb{5}} \qb{O} \right) 
+ Q_{\hbox{mix}} $$

where, according to Fig. 3, $ x $ is of the order of 4-5 and $ y $ of the order of
one%
\qfoot{In the framework of this reaction, the fact mentioned 
previously that $ Q_{\hbox{mix}} $ decreases when the temperature is 
increased becomes a simple consequence of the Le Chatelier principle for
equilibriums. Indeed, in order to oppose the temperature increase the
reaction must become less exothermic which means that the equilibrium is
shifted to the left.}%
.
Specific  details, and in particular the exact values of $ x $ and $ y $ are
anyway unimportant for the present discussion. The central point
is that the mixing creates a new molecular assemblage and that this
structural rearrangement basically results in an entropy {\it reduction}. 
Furthermore, it is of interest to observe that while the solution has
a greater cohesion than the initial compounds, it is also characterized
by a greater molecular agitation. The simultaneous occurrence of
higher cohesion and increased agitation is also observed in biological
and social systems. It is particularly spectacular in the
formation of swarms of locusts which we briefly discuss below.
\qpar

We said that  Fig. 3 is based on scattering experiments which permit
direct observation. However, it can also be justified by indirect arguments.
We mention that way of reasoning because it was the main approach in use
during several decades between 1905 and 1945. By measuring
specific physical variables for various compounds and by comparing 
these observations, physical chemists tried to gain a better understanding
of the processes taking place
at molecular level (see for instance Holmes 1913 or
Earp and Gladsstone 1935). In the present case, this kind of reasoning
can be illustrated as follows.
\qbu In order to see if our point about the roles of the two segments in
the ethanol molecule is correct, it is natural to examine what happens
when one of the segments becomes preponderant. When the hydrocarbon
chain becomes longer one would expect the alkane character of the
molecule to become more pronounced. This is indeed confirmed by
comparative observation for ascending members of the
alcohol family,
$ \qb{C}_n\qb{H}_{2n+2}\qb{O} $. For $ n=1,2,3 $ the alcohols are soluble
in water in all proportion which indicates the formation of strong links.
However, starting with $ n=5 $, the solubility is greatly reduced which shows
that the alkane character becomes predominant; 
for $ n=4,5,6 $ solubility is 0.11 g/ mole,
0.03g/mole and 0.06g/mole respectively. The solubility of the $ n=6 $ alcohol,
namely hexanol, is in fact of the same order of magnitude as the
solubility of the corresponding alcane, namely hexane, which is 0.001 g/mole.
The alkane-type behavior of high order alcohols is also observed at the
level of boiling temperatures. Whereas there is a huge difference of about
$ 220 \qdk $ between the boiling temperatures of methanol and ethane,
boiling temperatures of higher order alcohols tend asymptotically
toward those of the alkanes; thus for $ n=10 $ the difference is reduced
to less than $ 50 \qdk $. 
\qbu Conversely a behavior which becomes close to that of water is
observed when the role of the OH segment becomes predominant. 
Whereas for $ n \geq 2 $ all alcohols are soluble in hexane, methanol
($ n=1 $) is only slightly soluble at 0.12g/mole. 
Similarly the role of
the OH segment can be expected to be enhanced in di-alcohols or
tri-alcohols, that is to say molecules that contain two or three OH segments.
Thus glycerol, 
$$ \qb{H}_{\qb{2}}\!\!\! \qh{\qb{C}}{\qh{|}{\qb{OH}}} \ql  
\qh{\qb{C}}{\qh{|}{\qb{OH}}}\!\!\! \qb{H} \ \ \ql  
\qh{\qb{C}}{\qh{|}{\qb{OH}}}\!\!\! \qb{H}_{\qb{2}} $$

is soluble in  water in all proportions but is almost not
soluble in hexane                  
\qfoot{Although it has received great attention the question of solubility
does not seem to have been solved in a completely satisfactory way
from a theoretical perspective. Both the Hildebrand system and the more
recent Teas graph system are mainly empirical with little theoretical
justification. This, however, does not prevent them from being useful and
fairly accurate tools. Incidentally, it can be noted that an important
special case, namely the solubility of hydrocarbon polymers is well
accounted for by the Flory-Huggins theory based on combinatorial
counting. It does not apply to polar molecules however.}%
.
\qpar

Attempts to bridge the gap between statistical physics and socio-economic
phenomena usually come up against two difficulties.
The first one, which is not often mentioned, is
the ergodic hypothesis to which we come back later.
The second is the fact that the notions of entropy,
energy or temperature which are so central in physics have no obvious
counterpart in social phenomena%
\qfoot{As a matter of fact, this is not surprising
for these are macroscopic variables which are neither ``felt'' nor 
``known'' at molecular level.}%
.
That is why we carefully avoided using these notions. All the mechanisms
described in this first part can be transposed to biological or
social phenomena. This is the purpose of the second part of the paper.
\vskip 10mm

\centerline{\bf \Large Part II\quad Social phenomena in 
the light of physics}
\vskip 5mm

The main challenge in this part is to identify those (if any) social phenomena
which can be better understood in the light of the notions presented in
the first part. Needless to say, many kinds of social phenomena do not
fall into this category. For instance, cultural or gender studies
draw on notions which have no parallels in physics. There are however
many important socio-economic phenomena which can be interpreted
along the lines used in Part I. One can mention the following.
\qbu In the first part we emphasized the connection between interaction
strength and molecular rates of escape from a liquid. Any system whose
members are held together by some cohesion forces but may occasionally
escape from the system would provide a possible parallel. Table 1 
provides a number of examples. 



\begin{table}[htb]

 \small 
\centerline{\bf Table 1 \ Retention dynamics in various institutions}
\vskip 3mm
\hrule
\vskip 0.5mm
\hrule
\vskip 2mm

$$ \matrix{
\tvi 
\hbox{Institution} \hfill & 
\hbox{Intra-institutional bonds} \hfill & \hbox{Type of escape} \hfill \cr
\noalign{\hrule}
\qth 
\hbox{High school,} \hfill & 
\hbox{Links with students, teachers, professors;} \hfill  & \hbox{Dropout} \hfill \cr
\hbox{college, university} \hfill & 
\hbox{attraction of qualified jobs} \hfill & \hbox{} \hfill \cr
\hbox{} \hfill & \hbox{} \hfill & \hbox{} \hfill \cr
\hbox{Faith community,} \hfill & 
\hbox{Links with rest of congregation,} \hfill & 
\hbox{Decline in attendance} \hfill \cr
\hbox{religious order} \hfill & 
\hbox{common faith} \hfill & \hbox{} \hfill \cr
\hbox{} \hfill & \hbox{} \hfill & \hbox{} \hfill \cr
\hbox{Army} \hfill & 
\hbox{Patriotism, discipline, renumeration} \hfill & \hbox{Desertion} \hfill \cr
\hbox{} \hfill & \hbox{} \hfill & \hbox{} \hfill \cr
\hbox{Nation} \hfill & 
\hbox{Family ties,} \hfill & \hbox{Immigration} \hfill \cr
\hbox{} \hfill & \hbox{attraction of home country} \hfill & \hbox{} \hfill \cr
\hbox{} \hfill & \hbox{} \hfill & \hbox{} \hfill \cr
\hbox{Society} \hfill & 
\qtb 
\hbox{Family ties, links with friends} \hfill & \hbox{Suicide} \hfill \cr
\noalign{\hrule}
} $$

\vskip 1.5mm
Notes: The fact that one often observes a conjunction of
substantial high school dropout rates with
high suicide rates among teens seems to show that the interactions
which account for these effects overlap to some extent.
As an illustration, for the Oglala Sioux who live on
Pine Ridge Reservation, South Dakota, dropout  and
suicide rates among teens are 6 and 4 times higher respectively
than in the general population; for American Indians overall, the
dropout and teen suicide rates are 35 percent and 37 per 100,000, 
respectively 3 and 2.5 times higher than in the general population.
In the second column we attempted to list some of the possible
bonds that keep an individual attached to a given institution. This list,
however, is more based on common sense than on genuine
measurements. As a matter of fact, we do not yet know what is the
respective importance of these links. For instance, we know that family
ties are important in suicide, but we do not have a clear picture of the
respective role of short-range versus long-range ties. What makes 
reliable measurements difficult is the fact that the level of
exogenous shocks (which
represent thermal agitation) is usually time-dependent and has therefore
to be controlled for.
\qL
Sources: Reyhner (1992), Olson (2003), http://www.re-member.org
\vskip 2mm

\hrule
\vskip 0.5mm
\hrule

\normalsize

\end{table}


For instance, if the system is a college
or a university the retention rate of freshmen provides an indication
about the balance between group cohesion and centrifugal forces.
If the system is an army, the desertion or AWOL (Absent WithOut Leave) rates
provide a global estimate of the resultant of many forces such as for instance
patriotism, fear of being punished, conservation instinct, and so on. 
All the cases mentioned in table 1 would provide interesting testing
fields for the interpretation that we advocated. Unfortunately, for most
of them only scarce or fragmentary data are available. 
By a slight but natural extension it is possible to include suicide in the
present category. The likelihood of not committing suicide represents
a kind of retention rate. The main incentive for including suicide is the
fact that in this case at least there are numerous statistical data. Naturally,
suicide has been studied by sociologists for decades without any reference
to physics; one may therefore wonder what difference it makes to adopt
the present perspective. Instead of analyzing suicide statistics almost
indiscriminately, the present perspective leads us to focus on situations
where social ties are either very strong or very weak. By so doing we will
be in a better position to grasp the key mechanisms (as opposed to
incidental circumstances) of the phenomenon. The next section provides
an introduction to this approach.
\qbu The second physical phenomenon for which there are some natural
biological and social parallels is the condensation of a gas, that is to
say the transition from a state in which the molecules have low interactions
to a state where they form an entity characterized by a substantial
interaction and cohesion. Macromolecules, bacterias, protozoa, insects,
animals or humans in certain conditions display a tendency to 
self-aggregation. Instead of considering each of these cases as separate
it may help our understanding to look at them from a unified standpoint.
We briefly discuss two cases which belong to
this category of phenomena. 
\qbu The third physical phenomenon for which there is a natural 
sociological extension is the mixing of to liquids. Amalgamation of
different populations is a mechanism of fundamental importance.
Under that heading one can consider the amalgamation of populations
of peasants, merchants and craftsmen. Through the links of cooperation and
exchange that they establish, cohesion and productivity
are greatly enhanced. Another important mechanism
of amalgamation is the so-called melting pot mechanism by which
a group of immigrants becomes integrated. Again, one may ask what
benefit can be gained from considering these phenomena from the 
standpoint of statistical mechanics. In a physical solution the new bonds
between solute and solvent are established in a matter of seconds
if the solution is mixed up by an external device,
but it will take much longer if one has to rely on diffusion for the mixing
process. Similarly, the time scale required by the amalgamation process
very much depends upon the magnitude of the ``mixing''. 
It may take one or two generations in a city, but much longer in a
mountainous region where population density is low and
contacts are rare. This parallel shows that in order to understand the
dynamics of bond formation one must adopt an adequate time scale.
For urban integration, 50 years may be an acceptable time period, whereas
for low density regions two or three centuries would be more suitable.
In short, through the analogy with physical phenomena we get a better
understanding of how to set up the inquiry.

\qI{Suicide in a population with weak ties}

In the late 19th century there have been numerous studies about suicide
in all European countries.
The following references (arranged in chronological order) constitute a
select sample of the publications of that period, along with some more 
recent ones:
Boismont (1865),
LeRoy (1870), Cristau (1874), Morselli (1879),
Legoyt (1881), Masarick (1881), Nagle (1882),
Durkheim (1897), Krose (1906), Bayet (1922), Douglas (1967),
Baudelot et al. (1984). All these studies of course took advantage
of the fact that thanks to the development of census offices extensive
demographical statistics became available in all industrialized countries.
Among the aforementioned authors, the contribution of Emile Durkheim
stands out because, in contrast to most other authors, he was not 
interested in why individual people commit suicide but from the start
considered suicide as a social phenomenon. In the very first sections
of his book, he makes clear that to understand suicide one should 
examine the web of connections and affiliations each individual has
with the people around him. For Durkheim it is the failure of family,
church, community of neighbors to provide effective forces of social
integration which is at the heart of the problem. In short, Durkheim's
perspective is very close to the standpoint of statistical physics that
we presented in part I. Unfortunately, his message has been largely
discarded and forgotten,
to the point that nowadays most studies center on individual
psychological causes. 
\qpar

In support of his thesis Durkheim presents a great wealth of data 
for many different countries. However, for his argument to become really
compelling and conclusive one would need a way to measure the
strength of social ties in an objective and quantitative way. Instead
Durkheim relies on common sense and intuition with the result that his
proofs remain somewhat tautological. For instance, even if it is natural
to admit that bachelors have fewer social (and especially family) ties
than people who are married with several children, estimates based
on an objective criterion would be needed. Otherwise the observation that
suicide rates are higher among bachelors cannot be quite conclusive.
We must confess that our own methodology will have the same defect,
only to some extent mitigated by the fact that the situations that we
consider are so extreme that in order to make sense
our ``common sense'' estimates
need only to have the right order of  magnitude.
This is why we focus on situations characterized by low
levels of interaction. 
In the following we consider three situations of that kind.

\qA{People with schizophrenia}
Schizophrenia is a severe mental illness characterized by a 
variety of symptoms including lost of contact with reality and social
withdrawal. People with schizophrenia may avoid others or act
as though others do not exist; for example they may avoid eye contact
with others or may lack interest in participating in group activities.
Clearly this is a situation where interpersonal links are severely 
weakened. It turns out that suicide rates among people with schizophrenia
are 10 to 15 times higher than in the general population: a typical figure
is 200 per 100,000 as compared to 15 per 100,000 in the general population.

\qA{Inmates}
Persons who are arrested and jailed see
links with family, friends, colleagues or neighbors
suddenly severed. Of course, once in jail for some time, inmates are
likely to build new ties for instance with other inmates, guardians,
lawyers, chaplains or other persons who may assist them.
One would expect, therefore, that it is in the first days in jail that the
disaggregation of social ties is the most severely felt. This prediction
is matched by observation. Indeed, it
turns out that
suicide rates are particularly high during the first few days in jail.  A study
performed in 1986 about jail suicide in the US found that 51\% of the
suicides which occur in jail (as opposed to prison which in the US
designates facilities for stays of over one year) happen in the first
24 hours of incarceration. Thanks to
official data which are available on the Internet for New York State
(New York State 1998: Crime and Justice Annual Report, 
http://criminaljustice. state.ny.us)
we are able to compute an order of magnitude of 
the suicide rate in short-term detention facilities technically known
as ``lockups'' where detainees usually stay for less than 72 hours before
being transferred to county jails. The reasoning goes as follows.
\qL
On a single day of 1998 the average number of detainees in lockups was
473 (151 for New York City and 322 for upstate New York). Naturally,
these detainees were not the same throughout but this is irrelevant
for the present calculation. Over the whole year there were 6 suicides
(2 in New York City and 4 upstate) which gives a rate of 
$ 6/473 = 1268 \hbox{ per } 10^5 $. If the same calculation is done for each
year between 1990 and 1999 one gets an average suicide rate of $ 903 $
per $ 10^5 $. Because, a great majority of inmates are males,
this figure should be compared to the suicide rate of men
in the general population of New York State
which for the period 1990-1998 was $ 13.0 $ per $ 10^5 $.
The suicide rate in the first 6 days of detention was therefore 69 times higher,
This order of magnitude is consisted with results obtained
by other studies which analyzed
suicide rate in the first days of detention (table 2). As detainees form new
links in jail, the suicide rate progressively declines. In county jail where
inmates usually stay for periods of less than one year, the rate is about
10 times higher than in the general population. In state prisons, where 
inmates stay for periods of more than one year, the rate is almost the same
as in the general male population. 
\qpar

Table 2 provides also data for some other countries. These data do not 
distinguish
between short-term and long-term facilities. Most of the figures are 
between 100 and 200 which is consistent with the rates observed in
US county jails. 
\qpar



\begin{table}[htb]

 \small 
\centerline{\bf Table 2 \ Suicide rates among inmates}
\vskip 3mm
\hrule
\vskip 0.5mm
\hrule
\vskip 2mm

$$ \matrix{
\tvi 
&\hbox{Type of} \hfill & 
\hbox{Time elapsed} \hfill & \hbox{Location} \hfill & \hbox{Time} & 
\hbox{Suicide}  \cr
&\hbox{institution} \hfill & 
\hbox{since} \hfill & \hbox{} \hfill & \hbox{interval} & 
\hbox{rate}  \cr
\qtb
&\hbox{} \hfill & 
\hbox{incarceration} \ (T) \hfill & \hbox{} \hfill & \hbox{} & 
\hbox{[per 100,000]}  \cr
\noalign{\hrule}
\qth 
1&\hbox{Lockup} \hfill &  T< 72\  \hbox{hours} \hfill & 
\hbox{New York State} \hfill & 1990-1999 &  900 \cr
2&\hbox{Lockup} \hfill &  T< 72\ \hbox{hours}  \hfill & 
\hbox{South Dakota} \hfill &  1984 & 2975 \cr
3&\hbox{Jail} \hfill &   72\ \hbox{hours}< T< 1\ \hbox{year}\hfill & 
\hbox{Texas} \hfill &  1981& 137 \cr
4&\hbox{Jail} \hfill &  72\ \hbox{hours}< T< 1\ \hbox{year} \hfill & 
\hbox{South Carolina} \hfill &  1984&  166\cr
5&\hbox{Jail} \hfill &  72\ \hbox{hours}< T< 1\  \hbox{year} \hfill & 
\hbox{US} \hfill & 1986 & 107 \cr
6&\hbox{Jail} \hfill &  72\  \hbox{hours}< T< 1\  \hbox{year}  \hfill & 
\hbox{New York State} \hfill & 1986-1987 &  112\cr
7&\hbox{Prison} \hfill & 1\  \hbox{year} < T  \hfill & 
\hbox{US} \hfill & 1984-1993 & 21 \cr
\hbox{} \hfill &   \hfill &  \hbox{} \hfill &  &  \cr
8&\hbox{Not spec.} \hfill & \hbox{Not spec.}  \hfill & 
\hbox{Belgium} \hfill & 1872 &  190\cr
9&\hbox{Not spec.} \hfill &  \hbox{Not spec.}  \hfill & 
\hbox{England} \hfill & 1872 & 112 \cr
10&\hbox{Not spec.} \hfill &  \hbox{Not spec.}  \hfill & 
\hbox{Saxony} \hfill &  1872&  860\cr
11&\hbox{Not spec.} \hfill & \hbox{Not spec.}   \hfill & 
\hbox{Canada} \hfill & 1984-1992 & 125 \cr
12&\hbox{Not spec.} \hfill &  \hbox{Not spec.}  \hfill & 
\hbox{New Zealand} \hfill & 1988-2002 & 123 \cr
13&\hbox{Not spec.} \hfill &  \hbox{Not spec.}  \hfill & 
\hbox{England} \hfill & 1990-2000 & 112\cr
14&\hbox{Not spec.} \hfill & \hbox{Not spec.}   \hfill & 
\hbox{France} \hfill & 1991-1992 & 158 \cr
15&\hbox{Not spec.} \hfill &  \hbox{Not spec.}  \hfill & 
\hbox{Australia} \hfill & 1997-1999 &  175\cr
16&\hbox{Not spec.} \hfill & \hbox{Not spec.}   \hfill & 
\hbox{Canada} \hfill & 1997-2001 & 102 \cr
17&\hbox{Not spec.} \hfill &  \hbox{Not spec.}  \hfill & 
\hbox{Scotland} \hfill & 1997-2001 & 227 \cr
\qtb 
&\hbox{\bf Average (8-17)} \hfill &  \hbox{}  \hfill & 
\hbox{} \hfill &  &  \hbox{\bf 218}\cr
\noalign{\hrule}
} $$

\vskip 1.5mm
Notes: As a useful yardstick one can use the suicide rate among males
in the United States between 1979 and 1998 which was about 20
per 100,000.
Suicide rates of inmates are highly dependent upon the time 
they have spent in prison since their incarceration. A detailed study 
based on 339 suicides that occurred in the US in 1986 found that 51 percent
of the suicides occurred in the first 24 hours of incarceration. 
This observation is consistent with the interpretation of suicide as 
resulting from a severing of social ties. In the statistics published in other
countries than the United States, the time of incarceration is not specified.
However, since inmates incarcerated for less than one year 
are in greater number 
than those incarcerated for longer durations, one would expect
the former to predominate. Therefore it is not surprising 
that the order of magnitude of suicide rates is more or less the same
everywhere (one exception is Saxony).
\qL
Sources: 1: DCJS Report (tables 7,9,11); 2: Hayes et al. (p. 4); 3,4,5:
Hayes et al. (p. 52-53); 6: DCJS Report (table 1), Hayes et al. (table 2);
7: http://www.mces.org/Suicide\_ Prisons\_ Jails.html; 
8-10: Legoyt;
11: Correctional Service; 12: Corrections Department, 
http://www.corrections.govt.nz;
13: Her Majesty Prison Service; 14: Baron-Laforet, Bourgoin;
15-17: same as 12.

\vskip 2mm

\hrule
\vskip 0.5mm
\hrule

\normalsize

\end{table}


How accurate and reliable are the data given in table 2? This is certainly an
essential question. An official report (Hayes and Rowan 1998) 
found under-reporting of jail suicide in 1986
to be of the order 40 percent nationally,
but with great differences between states. Thus, in New York State no
under-reporting was identified (which is why we selected this state to
compute the previous estimate),
whereas in Alabama, Louisiana, Pennsylvania
or Tennessee under-reporting was over 50 percent. 
As there are no reasons for and indeed no mention of over-reporting 
we can at least be
assured that the figures which are made public provide 
trustworthy lower bounds.
\qpar

In conclusion, the phenomena of suicide among people with
schizophrenia or inmates seem to provide spectacular illustrations
of the effect of a weakening of social ties on suicide (escape) rates.
However these situations may be seen with good reason
as somewhat
artificial  in the sense that these groups are subject to illness or
special living conditions imposed from outside. This is why we now turn
to situations which can be considered as more ``natural'' in the sense
that they concern entire societies and occur  every time
``traditional'' societies come into contact with societies which are
technically more advanced. 

\qA{Traditional societies in situations of transition}

Every time the social framework of a society undergoes overwhelming 
changes there is a time of transition during which the old structures no
longer work or exist and those which are better adapted to the new situation
have not yet emerged. As a result, one would expect such periods 
of transition to be characterized by a low level of social interaction. 
When two liquids mix the time it takes for the
pattern of forces to rearrange and for the molecular structure to be reordered
is probably to be counted in microseconds at the molecular level and in
seconds at macroscopic level (provided the two liquids get mixed). In a 
society the transition may take decades. The figures in table 3 show that
these situations are characterized by a substantial increase in suicide 
rates up to levels which are 3 to 4 times higher than in stable societies.
These phenomena have many facets: familial, communal, 
demographic, economic, political, etc. In the rest of this section we 
focus our attention on the case of Micronesia for which one has fairly 
good statistical data.
\qpar


\begin{table}[htb]

 \small 
\centerline{\bf Table 3 \ Suicide rates in populations in a state of transition}
\vskip 3mm
\hrule
\vskip 0.5mm
\hrule
\vskip 2mm

$$ \matrix{
\tvi 
&\hbox{Population} \hfill & 
\hbox{Gender or age} \hfill & \hbox{Location} \hfill & \hbox{Time} & 
\hbox{Suicide}  \cr
&\hbox{} \hfill & 
\hbox{specification} \hfill & \hbox{} \hfill & \hbox{interval} & 
\hbox{rate}  \cr
\qtb
&\hbox{} \hfill & 
\hbox{}  \hfill & \hbox{} \hfill & \hbox{} & 
\hbox{[per 100,000]}  \cr
\noalign{\hrule}
\qth 
1&\hbox{Blackfoot} \hfill &   \hfill & 
\hbox{U.S.} \hfill &  1960-1969&  130\cr
2&\hbox{Cheyenne} \hfill &   \hfill & 
\hbox{U.S.} \hfill & 1960-1968 & 48 \cr
3&\hbox{Papago} \hfill &   \hfill & 
\hbox{U.S.} \hfill &  1960-1970& 100 \cr
4&\hbox{Indians} \hfill &   10-19 & 
\hbox{Canada} \hfill & 1986-1990 &  65\cr
5&\hbox{Natives} \hfill &  \hbox{Male} & 
\hbox{Micronesia} \hfill & 1975-1990 &  50 \cr
6&\hbox{Natives} \hfill &  \hbox{Male}, 15-24   & 
\hbox{Micronesia} \hfill &  1978-1987 & 129 \cr
\qtb 
7&\hbox{Natives} \hfill &  \hbox{Male}, 15-24   & 
\hbox{Chuuk Islands} \hfill &  1978-1987 & 200 \cr
\noalign{\hrule}
} $$

\vskip 1.5mm
Notes: As a matter of comparison the average suicide rate in the 
United States over the period 1979-1998 was 12.2 per 100,000 for both
genders, 19.5 per 100,000 for males and 20.7 for males aged 15-24.
The area considered in cases 5 and 6 comprises the Federal State of
Micronesia (Chuuk, Kosrae, Pohnpei, Yap), the Marshall Islands and
Palau. Blackfoot, Cheyenne and Papago are three tribes of American
Indians. Basically, either for North American Indians or in Micronesia
the suicide rate  of young adults is at least 4 times higher than in the
general population. 
\qL
Sources: 1-3: Lester; 4: http://www.hc-sc.gc.ca; 5-7: Rubinstein (2002)

\vskip 2mm

\hrule
\vskip 0.5mm
\hrule

\normalsize

\end{table}


Micronesia as it is defined by Donald Rubinstein (1994, 2002) from which 
we borrow most of the following information comprises the Marshall Islands,
the Carolinas which now form the Federated States of Micronesia and
the Northern Marianas. Most of these islands were occupied by Japan in
1914 and some of them were colonized by Japanese farmers.
After World War II they became American Territories until they acceded 
to some form of autonomy in the late 1980s. In the 1950s and 1960s nuclear 
tests and missile tests
were conducted at Bikini, Eniwetok and Kwajalein
located in the Marshall Islands. 
While alcohol consumption by the islanders had been regulated
or prohibited under Japanese rule, these restrictions were lifted in the
1960s. 
Moreover, thousands of US Peace Corps
Volunteers arrived, schools were built in every island and the economy
began to shift from a subsistence economy based on family gardening
and fishing to an economy based on imported products and wage
labor. Men played a central role in these two activities and their place
in society was more affected than women's activities which
centered around preparing food and taking care of the house and
children. Yet, on Fig. 4 it can be seen that suicide rates increased
for men as well as for women even if the latter did not exceed the level
observed in industrialized countries. 
\qpar

How can we integrate and interpret these various changes in terms of
social interaction? As we already noted, to 
do this in a satisfactory way would require either
quantitative information about the frequency and intensity
of interpersonal contacts and links, or a methodology (an equivalent to
infra-red spectroscopy) that would enable us to estimate interaction
strengths. However, we can make two observations which are
fairly revealing at least quantitatively. 
\qee{1)}  If interaction strength really
plays a key role in this phenomenon, one would predict that the
individuals who are most at risk are those who do not have strong
family connections. That is true for suicide in a general way, but one
would expect the effect to be much stronger in a situation 
where community bonds have been weakened.
In such a situation the transition
period between childhood, characterized by strong ties with parents,
and adulthood, characterized by strong ties with one's own wife and children,
is a particularly critical moment. In short, one would expect 
a high suicide rate for young adults. This is indeed what observation
shows (table 3). 
%
  \begin{figure}[htb]
    \centerline{\psfig{width=15cm,figure=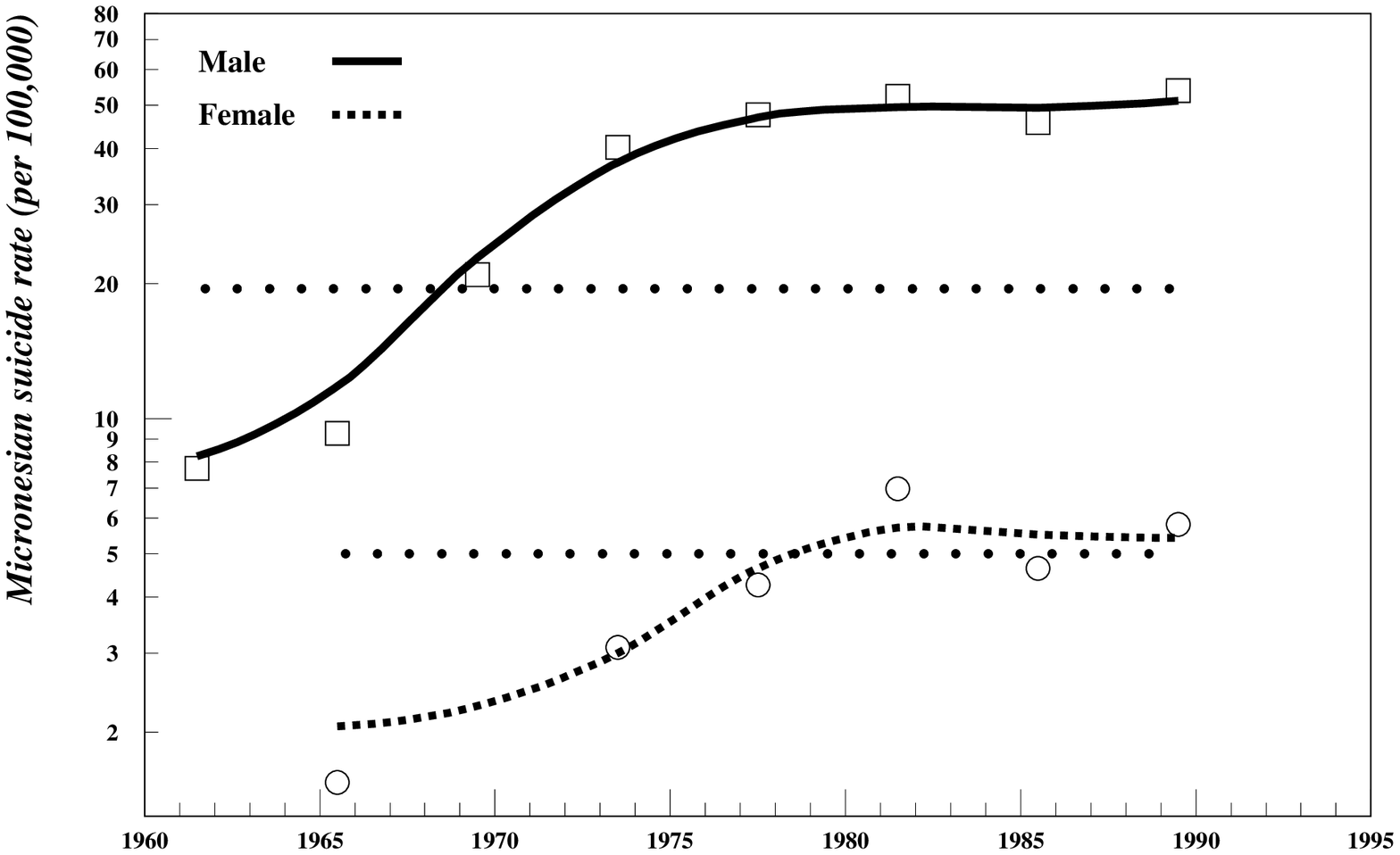}}
    {\bf Fig.4: Suicide rate in Micronesia}.
{\small The area covered by these data comprises the island of Palau,
the Federated States of Micronesia (which includes the Chuuk islands)
and the Marshall Islands. Concomitantly with the shift from a subsistence
economy to one based on imported food and wage labor, there has
been a huge increase in suicide rates for both men and women. The dotted
horizontal lines show the American
average suicide rates for men and women. The insert shows that it is 
the 15-24 age group which has by far the highest rate, a pattern which
strongly differs from the age pattern in industrialized countries where it
suicide rate increases with age.
It is of interest to note that in Micronesia the high school dropout rate
is around 40 percent that is to say four times more than in the United
States}.
{\small \it Source: Rubinstein (1994, 2002), Hezel (2001)}.
 \end{figure}
%
This effect is expected in any society in transition,
but in Micronesia it is to some extent amplified because of two local
factors. As it is (or at least was) considered taboo for a sexually mature
boy to sleep in the same house as his sisters, teenagers used to
move to community men's houses which they would share with extended
family relatives. But in many islands these community houses are no
longer kept up with the result that young men do not have places where
they can conveniently stay at this critical juncture. The second aggravating
circumstance is related to the custom of adoption. In Micronesia, adoption
was a widespread custom and a central pillar of family life. Like godfathers
or godmothers but in a much stronger sense, the adoptive parents 
were a major constituent of the enlarged family. However, this institution
has been imperiled both by the decay of traditional culture and by the
rapid population growth. With a large number of children in each family
there is much less incentive for adopting. The clear effect
of the decay of adoption was to reduce the links between teenagers and
adults. 
\qee{2)} Micronesia consists of more than 100 islands. As they have
been affected to different degrees by social change we are in a good
position for making comparative observations. Rubinstein (2002) mentions
that suicide is less frequent in
rural outer islands where traditional ways of life have to some extent
be maintained. For instance, in the central island of Chuuk the average
suicide rate of young males (15-24) was $ 207 $ per $ 10^5 $, whereas 500 
kilometers to the east in the Pohnpei and Kosrae islands it was only
$ 93 $ per $ 10^5 $. 
\qpar

At this point, we do not know if the factors that we have just described are
really the main mechanisms that account for the high suicide rates. In
order to confirm (or confute) these conjectures other similar cases must
be investigated in a comparative perspective.

\qI{Condensation phenomena}

Particularly abundant rain in an area of central or northern Africa may
result in a population of locusts which is in greater number than in ordinary
years. It seems that once the density of locusts reaches a critical level
a phase transition occurs which leads to the formation of swarms of locust
which may contains billions of insects and have a duration of several years.
In this process, locusts undergo slight physical changes (change in color,
development of a gland for the release of a specific pheromone) which lead
entomologists to describe them as a distinct subspecies (which they 
call a phase), namely {\it locust gregaria} as opposed to {\it locust solitaria}
(Faure 1932). In order to better understand the factors which determine
the transition from one phase to the other
Jacobus Faure raised population of locusts in adjacent cages.
It is worth
reporting how he describes a cage containing {\it locust solitaria}
in one half and {\it locust gregaria} in the other.
``Whereas an individual isolated
in one half of a cage leads a life of lazy idleness, its fellows in the
other compartment separated by a gauze partition only, live a life of
intense activity that could perhaps best be described as a frenzied effort
to escape from some relentless, inwardly pursuing force.'' This observation
suggests that the level of interaction and activity is greatly
increased in the swarm state.
\qpar 

After having established that it is the
crowding that leads to the formation of swarms, Faure went on to show
that inter-specific crowding has the same effect as intra-specific crowding,
In other words, the enhanced interaction that leads to swarm formation
seems to be unconnected to a particular species of locusts.
\qpar

The second example that we wish to mention of a transition 
analog to a condensation of a gas into a liquid is linked with the
phenomenon of territorial conquest by clans of nomads that occurred
repeatedly in central Asia. At least 5 episodes have been recorded by
historians between the first and 17th century. In their ``gaseous'' state
the clans wander more or less randomly over the vast land expanses of
the steppe and have only minimal interaction with one another. 
Then, all of a sudden for no obvious reason, the various clans will
gather around a leader whether it is Attila, Genghis Khan or Tamerlane.
On such episodes our historical information is mostly limited to 
anecdotal evidence.
We largely ignore the reasons which trigger these events%
\qfoot{Demographic factors may have played an important role.
Historians are not very informative on this aspect, but it is suggested almost
as an evidence by Machiavelli (1469-1527), the Italian statesman and 
historian who lived in a time when the remembrance
of Tamerlane (1370-1405) was probably still alive. In 1517 he wrote in
his {\it Discourses on Livy}:
``The other kind of war is when an entire people with all their 
families goes to seek a new seat in a new province.
These people go out from their countries driven 
by necessity; and the necessity arises from famine, or war, and oppression, 
which in their own country is experienced by them.
These people are
almost all from the country of Scythia, a cold and poor place, where,
because there were a great number of men and the country of a kind
which was unable to feed them, they are forced to go out, having many
things which drive them out and none to retain them.'' (adapted from
the English translation of 1675, 
http://www.constitution.org/mac/disclivy\_ .htm).}
; 
however, once started, it may have worked like a chain reaction
in the sense that
any battle won by the new leader strengthened and broadened his 
force of attraction and brought clans closer together. The physical analog
would consist in following an isotherm on a (volume per mole, pressure)
phase transition diagram; as one moves 
from the region of high volume per mole (i.e. low density) to lower
volume per mole, one first sees some droplets of liquid form, and then, 
progressively, an ever larger proportion of gas becomes liquid
until eventually all the gas is liquid.
If this mechanism is correct, it means that the episodes of territorial 
conquest experienced by the nomadic tribes of Central Asia have
been triggered by a major population increase. 
\qpar

What was the threshold density of population, in other words
what is the upper density limit for 
nomadic people living on
their livestock and occasional hunting. This is certainly an important
parameter and in order to find its
order of magnitude the most reliable procedure is probably to look at
present day population density in regions
where this mode of subsistence is still in effect. 
The following figures are for 1994 (i) The
density of the vast area which comprises the Republic of
Mongolia, Inner Mongolia (part of
China) and Buryatia (a part of historical Mongolia now located in Russia)
was 8.4 inhabitants per square kilometer. 
(ii) The density of Xinjiang in western China was
9.7 (iii) The density of the Chinese province of Qinhai which is
located between Xinjiang and Tibet was
6.5. In all these cases the density of the grazing livestock
was between two to three times the human density. To sum up, one can 
conclude that the upper density limit was around
8 people per square kilometer with about 2.5 head of livestock for each
person. It should be noted that this limit is an upper bound but that
the actual critical population density may have been somewhat lower.
Once more population data about the previous historical episodes
become available, it will be possible to say if our conjecture that
they were driven by a common mechanism is indeed confirmed.

\qI{Multicomponent solutions and the melting pot mechanism}

The integration of new immigrants is easier in large cities than
in rural towns. Why? It is of course easy to provide a number of
``anthropological" reasons such as better economic opportunities or
the fact that the inhabitants of large cities are more used to 
the presence of various immigrants than the people in rural towns.
Here we want to see whether our parallel with solutions
can tell us something about this question. Two observations can
be made in that respect.
\qbu Roughly speaking, solubility requires that the molecules of the
solute come in between the molecules of solvent and that new
links are established between the former and the latter. For this to occur
the interactions between solvent molecules should not be too strong
whereas the solvent-solute interactions should be as strong as possible.
Translated into sociological language, the fact that integration 
is easier in cities means that the interactions in cities (solvent) are 
weaker than in towns or that immigrant-city interactions are greater
than immigrant-town interactions ( both can conditions can be fulfilled
simultaneously). 
Intuitively, these conditions seem to agree with common sense,
but once again we are hampered by our inability to {\it measure} 
the strength of interactions.
\qbu The previous discussion is not completely realistic because
it considers a binary solution whereas in social situations
there are several components. Although
physical data handbooks contain less
information about multicomponent solutions than about binary solutions,
one aspect appears very clearly: multicomponent solutions are much
less selective than binary components. What do we mean by the expression
``less selective''? The heat of mixing curves for water-alcohol (Fig.1b)
resemble resonance curves; the fact that there is a sharp fall on
both sides shows that energetically the solvation is much less 
favorable as soon as one leaves the peak region; 
in other words the solubility is fairly selective. In contrast, for
a multicomponent solution the heat of mixing curves are almost flat
which shows that the solvation has a low selectivity. Intuitively, this
is easy to understand for if the solution already comprises 
various molecules characterized by different kinds of bonds, it will be
easy for any new molecule to link itself to one of them.
This property of multicomponent solutions is illustrated in a more
quantitative way in the following table.


\begin{table}[htb]

 \small 
\centerline{\bf Table 4 \ Relative selectivity of solvation according to
number of components}
\vskip 3mm
\hrule
\vskip 0.5mm
\hrule
\vskip 2mm

$$ \matrix{
\tvi 
&\hbox{2 components}  & \hbox{3 components}  & \hbox{5 components}   \cr
\noalign{\hrule}
\tvi
\hbox{Coeff. of var. of Q / Coeff. of var. of proportion} \hfill &  77\% & 
24\% &   12\% \cr
\noalign{\hrule}
} $$

\vskip 1.5mm
Notes: The table tells us that a
solution with several components is less ``selective''
than a binary solution. The percentages
give the ratio of the coefficient of variation (i.e. standard
deviation divided by mean) of the 
heats of mixing, $ Q $, relative to the coefficient of 
variation of changes in the proportion
of one of the components. As an illustration,
for water + ethanol (considered in Fig.1b)
the ratio is for instance equal to 0.98.
The 2-component figure is an average over three cases: water + methanol,
water + ethanol, ethanol + toluene; the 3-component figure is the average of
the two following cases:
benzene + cyclohexane + hexane (proportion changes refer
to hexane), benzene + cyclohexane + n-heptane (proportion changes refer
to heptane); the 5-component case is: benzene + hexane + toluene 
+ cyclohexane + heptane (proportion changes refer to heptane).
\qL
Source:  Landolt-B\"ornstein (1976, p. 536-540).

\vskip 2mm

\hrule
\vskip 0.5mm
\hrule

\normalsize

\end{table}


The more components there are, the less changes in the molar proportion of
one of the components perturbed the solution, as reflected in the fact
that the heat of mixing becomes almost independent of the proportion.

\qI{Conclusion}

Although I did not propose any model, time is not yet ripe for that,
I hope this paper will help us to see a number of socio-economic
phenomena in a more unified and less anthropocentric way.
The potential usefulness of the parallels developed in this paper
is that it gives us the incentive to compare phenomena which
at first sight seem to have little in common.
In this concluding section I would like firstly to discuss the question
of ergodicity, an important theoretical issue on which depends the
applicability of statistical mechanics, and secondly to suggest an
agenda for future research.

\qA{Does the ergodicity hypothesis hold for socio-economic systems?}

The success of statistical mechanics is entirely based on the fact
that ensemble averages can be identified with time averages. 
On the theoretical side we compute the most probable 
configurations of the system on the basis of a collection of similar
systems characterized by the same initial conditions and macroscopic
constraints. The classical example is a gas contained in a container
in a state of equilibrium. Strictly speaking,
the probability of finding all the molecules in one half of the
container is not null, but it is overwhelmingly smaller than the probability of
the situations where the molecules are uniformly distributed (except for
small random fluctuations). The assumption that the time
the system under observation spends 
in each macrostate is proportional to the probability of this state rests on
the hypothesis not only that the system randomly explores all
accessible microstates, but that it explores them ``quickly enough''.
As we have seen, for molecules in a gas or in a liquid the typical duration
of a given configuration is of the order of one picosecond which means
that within the time it takes to make a measurement the system
explores over $ 10^{12} $ configurations. 
\qpar

No matter how we define the
configuration space, it is obvious that it will be explored much more slowly
in the case of socio-economic systems. For instance, on stock markets
probably one of the economic systems with the highest
transition rate, there are on average less than 10 transactions per second
even for the most heavily traded stocks. For other socio-economic
systems the number of transitions may be smaller by several orders of
magnitude. This has at least two consequences. (i) The time it may take 
for equilibrium to be reached may be large compared to the time
scale of human observation. (ii) Consequently, there is a substantial 
probability of seeing the system in some metastable state rather than
in its ``true'' equilibrium state. As a matter of fact, this problem 
is not specific to social systems; it also exists
for some physical systems such as selenium, sulfur or tin which have
different allotropic forms. For instance the transition from
white tin to gray tin is supposed to occur at $ 13 \qd $, but it may take
centuries for a plate of white tin to decompose into gray tin
even at temperatures as low as $ -18 \qd $ (Kariya et al. 2000,
http://www.natmus.dk). Most of the tools developed in statistical mechanics 
are not well suited to such systems.

\qA{An agenda for future research}
One may wonder whether the three manifestations of suicide that we
examined can be accounted for by the same mechanism in spite of
the fact that they correspond to very different time scales ranging
from a few hours to several decades. To try to build a model at this point
would probably be premature. If the model ``explains'' observed 
suicide rates in terms of social bonds that we cannot estimate in
an independent way, this would be no more than a form of circular
reasoning. This shows that one of our most urgent tasks is to develop
methods for estimating the strength of social interaction. That this
objective has so far been largely ignored by sociologists
may seem surprising. How, for instance, is it possible to understand
revolutions which basically consist in a rearrangement of
social networks, if one has no real means for assessing the strength of
social ties?
\qpar

What approach can we think of for that purpose? In physics, experimental
methods for measuring the strength of molecular bonds 
involve infra-red spectroscopy, ultra-sonic spectrography or
X-ray/neutron scattering. 
The three methods are similar in their principle. A wave
is sent through the medium and the various ways in which it is affected
are recorded and used to probe various characteristics
of the medium. 
The first method is slightly different from the two others in the sense that
it relies on
the absorption which occurs when the frequency of the 
source coincides with the stretching or vibration modes of the 
molecular structure. So far we do not know much about the eigenfrequencies
(if any) of socio-economic systems which means that for the time being this
is not a straightforward approach. The two other methods have a broader
applicability. For instance, in ultra-sonic spectrography an
ultra-sonic wave is transmitted through the medium, its velocity and attenuation
are recorded, and from these measurements one may derive 
many properties of the medium, for instance its density, the size
of emulsion droplets or the existence of a temperature gradient.
At this point, an
important requirement needs to be emphasized which is
crucial if one wants to extend this approach to the social sciences.
In order to be able to use light or sound waves as probes, one must
already know how  specific characteristics of the medium are likely to affect
these signals.
In other words, it is only once we have gained some 
kind of understanding (even if it is only an empirical understanding) 
of how a given signal is affected by a society that
we can use it as a probe. But once this condition is fulfilled, this probe 
will allow more detailed and
systematic explorations. In short, this is a kind of cumulative process where
any new knowledge about social interactions is used as a springboard 
for further explorations. 
\qpar

There are many social signals which should enable us to transpose
the previous approach to the social sciences. For instance, the way
an epidemic propagates in a society can reveal a lot about the
interactions that take place in a population.
As an example, one can mention the fact
that if two population groups show
highly different levels of prevalence for a sexually
transmissible disease such as gonorrhea or chlamydia, one can be almost
sure that the two groups have little interactions through marriage or
non-marital sexual contacts. Such assessments can be checked 
by confronting them against inter-marriage rates; alternatively, they can
replace such statistics in those countries which for some reason do not
record that kind of data.
Naturally, this example concerns
only one particular aspect of social interaction. One would have to develop
similar approaches for other aspects as well. The
transmission of rumors, innovations or new fashions can provide
other possible probes.
\qpar

In this paper, I tried to convince the reader of the key role of
interaction strength in social phenomena. I am convinced that
once we get a clearer picture of this factor
many socio-economic phenomena will become more transparent and,
to some extent, more predictable.

\vfill \eject

{\large \bf References}
\vskip 5mm

\qparr                                                                   
Baron-Laforet (S.) 1999: Rep\'erage du suicide en prison et \'el\'ements
contextuels. 
http://psydoc-fr.broca.inserm.fr

\qparr
Baudelot (C.), Establet (R.) 1984: Durkheim et le suicide. 
Presses Universitaires de France. Paris.

\qparr
Baumert (J.), Gutt(C.), Tse (J.S.), Johnson (M.R.), Press (W.) 2003: Localized
guest vibrations and lattice dynamics of methane hydrate. Oral presentation
at the 3rd European Conference on Neutron Scattering. Session B. 
Montpellier, September 2003.

\qparr                                                                   
Bayet (A.) 1922: Le suicide et la morale. 
F\'elix Alcan. Paris.

\qparr
Boismont (A.B.) 1865: Du suicide et de la folie suicide. 
Germer Bailli\`ere. Paris.

\qparr
Bose (E.) 1907: Resultate kalorimetrischer Studien.
Zeitschrift f\"ur Physikalische Chemie 58, 585-624.

\qparr
Bourgoin (N.) 1999: Le suicide en milieu carc\'eral.
Population, May-June, 609-625.

\qparr                                                                   
Correctional Service of Canada 1992: Violence and suicide in Canadian
institutions: some recent statistics. Forum 4,3,1-5.

\qparr                                                                   
Corrections Department of New Zealand 2003: Inmate deaths in custody.
http://www.corrections.govt.nz

\qparr
Cristau (C.-A.) 1874: Du suicide dans l'arm\'ee. Thesis. Paris.

\qparr                                                                   
DCJ Report. New York State Division of Criminal Justice Services 1999:
Crime and justice annual report 
(available at http://criminaljustice.state.ny.us)

\qparr                                                                   
Dill (K.A.), Bromberg (S.) 2003: Molecular driving forces: Statistical 
thermodynamics in chemistry and biology.
Garland Science, New York.

\qparr                                                                   
Dixit (S.), Crain (J.), Poon (W.C.K.), Finney (J.L.), Soper (A.K.) 2002:
Molecular segregation observed in a concentrated alcohol-water
solution.
Nature 416,6883,829-832.

\qparr                                                                   
Douglas (J.) 1967: The social meanings of suicide. 
Princeton University Press. Princeton.

\qparr                                                                   
Durkheim (E.) 1897: Le suicide: \'etude de sociologie. 
F\'elix Alcan. Paris.

\qparr                                                                   
Earp (D.P.), Glasstone (S.) 1935: Dielectric polarization and 
molecular-compound formation in solution.
Journal of the Chemical Society 1709-1723.

Faure (J.C.) 1932: The phases of locusts.
Bulletin of Entomological Research 23,293-428.

\qparr                                                                   
Guo (J.H.), Luo (Y.), Augustsson (A.), Kashtanov (S.), Rubensson (J.E.),
Shuk (D.K.), \AA gren (H.), Nordoven (J.) 2003: Molecular structure
of alcohol-water mixtures.
Physical Review Letters 91,15,157407.

\qparr                                                                   
Hayes (L.M.), Rowan (J.R.) 1988: National study of jail suicides: seven
years later.
National Center on Institutions and Alternatives.

\qparr                                                                   
Her Majesty Prison Service 2001: Prison suicides. London.

\qparr                                                                   
Hezel (F.X.) 1987: Truk [called Chuuk after 1990] suicide epidemic and
social change.
Human Organization 48,283-296.

\qparr
Hezel (F.X.) 2001: Competition drives school improvement.
Pacific Magazine and Island Business (May).

\qparr                                                                   
Holmes (J) 1913: Contributions to the theory of solution: The 
intermiscibility of liquids.
Journal of the Chemical Society 2147-2166.

\qparr                                                                   
Israelachvili (J.), Wenerstr\"om (H.) 1996: Role of hydration and water
structure in biological and colloidal interactions.
Nature 379, 18 January, 219-225.

\qparr                                                                   
Kariya (Y.), Gagg (C.), Plumbridge (W.J.) 2000: The pest in lead free
[i.e. pure tin] solders. 
Soldering and Surface Mount Technology 13,1,39-40.

\qparr                                                                   
Krose (H.A.) 1906: Der Selbstmord im 19. Jahrhundert.
Herdersche Verlag. Friburg.

\qparr
Landholt-B\"ornstein edited by K. Sch\"afer 1976: Numerical data and
functional relationships in science and technology. Group IV, Vol. 2:
Heats of mixing and solution. Springer. Berlin.

\qparr                                                                   
Legoyt (A.) 1881: Le suicide ancien et moderne: \'etude historique,
philosophique et statistique.
A. Droin. Paris.

\qparr                                                                   
LeRoy (E.) 1870: Etude sur le suicide et les maladies mentales dans le
d\'epartement de Seine-et-Marne. Victor Masson. Paris.

\qparr                                                                   
Lester (D.) 1997: Suicide in American Indians.
Nova Science. New York.

\qparr                                                                   
Lide (D.R.) ed. 2001: CRC Handbook of chemistry and physics.
CRC Press. Cleveland.

\qparr                                                                   
Machiavelli (N.) 1986: The discourses. 
Viking Penguin. New York.

\qparr                                                                   
Masaryck (T.G.) 1881: Der Selbstmord as sociale Massenerscheinung 
der modernen Civilisation. 
Carl Konegen. Vienna.

\qparr                                                                   
Moelwyn-Hughes (E.A.) 1961: Physical chemistry.
Pergamon Press. Oxford.

\qparr                                                                   
Morselli (E.) 1879: Il suicido saggio di statistica morale comparata.
Fratelli Dumolard. Milan.

\qparr
Nagle (J.T.) 1882: Suicides in New York City during the 11 years ending
Dec. 31, 1880. Riverside Press. Cambridge (Ma.).

\qparr
Olson (L.M.) 2003: Suicide in American Indians. University of Utah.

\qparr
Reyhner (J.) 1992: American Indians out of school: A review of 
school-based causes and solutions. 
Journal of American Indian Education 31,2.

\qparr                                                                
Rubinstein (D.H.) 2002: Youth suicide and social change in Micronesia.
Occasional Paper (Kagoshima University) 36,33-41.

\qparr
Schmid (R.) 2001: Recent advances in the description of the structure
of water, the hydrophobic effect, and the like-dissolves-like rule.
Chemical Monthly 132,11,1295-1326.

\qparr                                                                   
Yalkowsky (S.H.) 1999: Solubility and solubilization in aqueous media.
Oxford University Press. New York.

\end{document}